# Quantized Indexing: Beyond Arithmetic Coding

Ratko V. Tomic[*]
1stWorks Corporation[†]

Quantized Indexing[‡] is a fast and space-efficient form of *enumerative coding*[1-11], the most "*desirable*"[4,6] among asymptotically optimal universal source coding algorithms. The present advance in enumerative coding is similar to that made by arithmetic coding[17-22] with respect to its unlimited precision predecessor, Elias coding[12-16]. The arithmetic precision, execution time, table sizes and coding delay are all reduced by a factor $O(n)$ at a redundancy below $\log(e)/2^{g-1}$ bits/symbol (for *n* input symbols and a *g*-bit QI precision). Due to its *tighter enumeration*, the QI redundancy is below that of arithmetic coding (which can be derived[23] as a lower accuracy approximation of QI). The relative compression gain vanishes in large *n* and in high entropy limits and increases for shorter outputs and for less predictable data. QI is *significantly faster* than the fastest arithmetic coders[22], from *factor* 6 in high entropy to over 100 in low entropy limit ('typically' 10-20 times faster). These speedups are result of using only 3 adds, 1 shift and 2 array lookups (all in 32 bit precision; $\alpha=2$[§]) per less probable symbol and *no coding operations* for the most probable symbol. Further, the exact enumeration algorithm is sharpened and its lattice walks formulation is generalized. A new numeric type with a broader applicability, *sliding window integer,* is introduced.

## Evolution of Enumerative Coding

Enumerative coding (**EC**) appeared in entropy coding literature in 1960s[2,3], although the algorithm has been known in mathematics since 19$^{th}$ century[**]. It was formulated as a general entropy coding algorithm by Cover[1] in 1973 who named it "enumerative coding". A larger significance of EC – as a benchmark for modeling flexibility and resilience – was identified that same year by Davisson[4] who used it as the master algorithm[††] for the best type of universal codes theoretically possible, the minimax universal codes. Despite its theoretical appeal, EC was impractical as a general coding algorithm because it required arithmetic precision of the size of output. Although Cover had suggested a path toward solving the EC precision problem – a "*suitable function of low complexity*" providing an "*integer upper bound*" for the enumerative addends ([1] p. 76) – the "*suitable function*" remained elusive. This led Rissanen to comment few years later that this "*conveniently calculated upper bound, will not work at all*", because satisfying "*the

---

[*] Chief Scientist and CTO, 1stWorks Corp., rvtomic@1stworks.com, Lexington MA
[†] 30 Noon Hill Ave., Norfolk MA 02056, USA, tel: (508) 541-6781, http://www.1stWorks.com
[‡] For extended reference, papers, tech. reports and code: http://www.1stWorks.com/ref/qi.htm
[§] For alphabet $\alpha > 2$, the complexity scales as $C \cdot \log(\alpha)$. An alternative algorithm[24], more suitable for low entropy sources, scales as $C \cdot \log(h)$, where *h* is entropy/symbol, but this variant has a larger multiplier *C*.
[**] Donald Knuth attributes the **combinatorial number system**, which is a mathematical form of EC, to **Ernesto Pascal**, *Giornale di Matematiche* **25** (1887), 45-49 (cf. *The Art of Computer Programming* Vol. 4, 7.2.1.3, p. 6; preprint at: http://www-cs-faculty.stanford.edu/~knuth/fasc3a.ps.gz ).
[††] Arithmetic coder can also code the 'universal way', albeit less accurately and much more slowly than QI.

*invertibility requirement together with the decodability condition becomes a major difficulty*" ([18] p. 49)[*]. Instead, Rissanen found[17,18] that by first approximating[†] the exact enumerative addends to within $O(\log(n)/n)$ bits/symbol, the *subsequent r-bit truncation* of these approximate addends was decodable and the truncation resulted in an additional redundancy of at most $\log(e)/2^{r-2}$ bits/symbol[15,18]. The Rissanen's algorithm[17] and its descendants, which became known as **arithmetic coding** (**AC**), had thus solved the precision problem, but *only* for the already approximate enumeration and *only* with further adverse side-effects[‡].

The precision problem of the *exact* enumeration had resisted solution, despite various attempts[5-11] toward practical EC over the past three decades[§]. The most common[5-9] remedy was to code in smaller blocks (of fixed or variable size) and then try to reduce the resulting overhead arising from the increased number of small symbol counts and fractional bit losses at block boundaries. In [5],[7] variable input blocks were used with codes which don't require that block symbol counts be sent separately from index. Even these partial solutions, while not competitive against AC overall, have demonstrated distinct advantages of EC in certain areas, especially for coding of less predictable sequences (e.g. video codecs[8] and lossless coding of complex images[9]). A direct attack on the arithmetic performance problem was proposed recently by Ryabko[10] based on the Schönhage-Strassen fast multiplication algorithm. Although the asymptotic performance appeared much improved in theory, no practical algorithm has emerged from this effort as yet. Within the specialized domain of constrained coding[**], Immink[11] proposed a method based on floating point approximation of addends. Although the decodability problem of truncated EC addends encountered by Rissanen emerged here as well, Immink suggested a form of brute force exclusion of the non-decodable strings[††].

Quantized indexing (QI) is a direct and optimal solution for the precision problem of exact message enumeration. Unlike the double approximation of AC, QI performs only the precision reduction, which is the sole approximation truly necessary, and it selects the truncation with the lowest redundancy possible at a given precision. To prepare the natural setting for QI, we will first reformulate the conventional EC[1] as follows: (*a*) construct an important and general algorithmic component of EC previously overlooked, (b) extend the lattice walks formulation[5] of EC, (c) reverse the ranking convention[1] from lexicographic (***lex***) to co-lexicographic[26] (***colex***).

---

[*] In our formulation of EC, these two Rissanen's requirements correspond to our eqs. (20) and (8).
[†] The Rissanen's approximation of exact enumeration consists of applying Stirling approximation to the exact enumerative addends (binomials), then further approximating these by dropping the sqrt() and the $O(1+1/n)$ factors (cf. [17] eq. (5), the addends $\Phi$ which majorize the EC binomials), resulting in total excess bits > $\log(2\pi n pq)/2$ before any precision reduction. This AC redundancy for $n < \infty$ is absent in QI.
[‡] The most unfavorable among these are the large coding speed penalty and the hardwiring of the 'probabilty of single next symbol' modeling interface bottleneck (cf. [23], pp. 19-25, 30-31).
[§] For a pedagogical survey of EC algorithms (with pseudocode) see [8] and for related combinatorics [26].
[**] Constrained coding is used extensively for low level recording media codes. Even the conventional EC, which has been used in this field since 1950s, is advantageous here over AC and prefix codes due to small data blocks, intricate constraints and feasibility of direct brute force exhaustive enumerations.
[††] Although Immink doesn't demonstrate any collision resolution algorithm, the brute force removal of collisions should be attainable in this domain by virtue of much smaller combinatorial space and regular bit patterns for the colliding codes (easily identifiable and known upfront, being the constraints).

# Enumeration via Counting of Lattice Paths

The basic idea of EC is to map sets of equiprobable messages[*] into sets of combinatorial objects, *enumerative classes*, for which there are efficient un/ranking[†] procedures. To encode a message $\mathcal{M}$, the encoder selects enumerative class $E(\mathcal{M})$ and maps the message into an object $X(\mathcal{M}) \in E(\mathcal{M})$, then it computes the index $I(X \in E)$ of the object within its class (ranking). The encoder output is the identifier for the class, the *class tag* $C(E)$, plus the object index $I(X \in E)$. The decoder uses the class tag $C$ to identify $E$ and the index $I$ to reconstruct the object $X$ via unranking of $I$ within $E$, then it maps $X$ back into $\mathcal{M}$.

Following Schalkwijk[5], our combinatorial setting is an *R*-dimensional integer lattice of points $M = (x_0, x_1, ..., x_{R-1})$ with integer coordinates $x_r \equiv x_r(M)$, $r = 0..R-1$. A *step* along the *r*-th dimension is a unit vector $<r> \equiv (0,...,0,1,0,...,0)$ where the single 1 has *r* zeros to its left. The inverse function $r = d(<r>)$ returns dimension index of a vector. To "make a step" $<r>$ from point $M'$, we add the two vectors: $M' + <r> = (x_0,..., x_r+1,..., x_{R-1}) \equiv M$ and call the point $M' = M - <r>$ a *predecessor* of point $M$. Our basic 'combinatorial object' is a *lattice path* $T_n$, which is a sequence of points $T_n \equiv M_0, M_1, ... M_n$, where $M_0 \equiv (0,0,...)$, $M_i \equiv M_{i-1} + <r_i>$ for $i = 1..n$ and $<r_i>$ is a sequence of *n* steps defining the *n*-step path $T_n$ (thus we also write paths as a sequence of their steps: $T_n \equiv r_1 r_2...r_n$). We denote "path ending at $M$" as $T(M)$, "set of all paths ending at $M$" as $E(M)$, "count of all paths ending at $M$" as $N(M) \equiv |E(M)|$ and "extension of path $T$ by step $<r>$" as $T+<r>$. A symbol sequence $S_n \equiv a_1 a_2... a_n$ where $0 \leq a_i < \alpha \equiv R$ maps to a lattice path $T_n$ via $r_i \equiv a_i$, $i = 1..n$ (the input symbol $a_i$ results in the *i*-th step of lattice path along $a_i$-th dimension). Using *Iverson's selector*[25] $[B]$ (where $B$ is a boolean expression and $[true] = 1$, $[false] = 0$), the translations $S_n \leftrightarrow T_n$ are:

$$S_n \text{ to path } T_n \implies x_r(M_i) = \sum_{j=1}^{i} [a_j = r] \tag{1}$$

$$M_i = M_{i-1} + <a_i> = \sum_{j=1}^{i} <a_j> \tag{2}$$

$$T_n \text{ to string } S_n \implies a_i = d(M_i - M_{i-1}) \tag{3}$$

$$n = \sum_{r=0}^{R-1} x_r(M_n) \tag{4}$$

These mappings allow us to use paths $T_n$ and sequences $S_n$ interchangeably. Eq. (1) implies that the count of symbol $r$ in $S_n$ is $x_r(M_n)$. We call a point $M_n$ solving eq. (4) for given $n$, the *n*-step point $M_n$ and the set of all such points the *n*-step **front** $F_n$. Hence $F_n$ is the set of all points 'reachable' in exactly *n* steps (in our lattice step metrics such set is on a hyperplane eq. (4), while in the Euclidean metrics it would be on a surface of a sphere).

---

[*] The job of EC modeling engine is to map the input sequence into 'equiprobable messages' (usually segments of the input sequence) and select the enumerative classes ([23] pp. 26-35, on EC modeling).
[†] *Ranking* of a set $S$ is a reversible mapping $I(x)$ which maps each $x \in S$ into unique integer $I(x)$, called *index* of $x$: $0 \leq I(x) < |S|$. *Unranking* is the inverse mapping (of index into set elements): $I(x) \to x$.

To represent often encountered constraints, such as *mixed radix* codes where $0 \leq a_i < R_i$ and $R \equiv \text{Max}\{R_i: i=1..n\}$, we introduce **boolean constraint evaluator $M{:}r$** which is *true* (*false*) if step $<r>$ is allowed (disallowed) *from* point $M^*$. For mixed radix codes with given radices $R_i$, the constraint evaluator for $(i+1)$-st step is: $[M_i{:}r] = [\, r < R_{i+1}]$.

The key building block for our path ranking algorithm is the count of paths arriving at $M$, $N(M) \equiv |E(M)|^{\dagger}$. In the bottom up approach we want to express the $i$-step path count $N(M_i)$ in terms of $(i-1)$-step path counts $N(M_{i-1})$. Note that any path reaching $M_i$ arrives from some $(i-1)$-step point $M_{i-1}(r)$ via some last step $<r>$, thus $M_{i-1}(r) = M_i - <r>$. Each point $M_{i-1}(r)$ contributes its $N(M_{i-1}(r))$ paths to $N(M_i)$, $r = 0..R-1$ (if step $<r>$ is allowed), hence:

$$N(M_i) = \sum_{r=0}^{R-1} [M_{i-1}(r) : r] \cdot N(M_{i-1}(r)), \quad i = 1,2,\ldots \qquad (5)$$

$$N(M_0) \equiv 1, \quad M_0 \equiv (0,0,\ldots,0) \qquad (6)$$

A lattice for binary input ($R=2$) with fronts $F_1..F_8$ is depicted[‡] below. The lattice points are $(x_0, x_1) \equiv (x, y)$, with steps: $<0> = (1,0)$ and $<1> = (0,1)$. The path counts are computed via (5), advancing from $F_1$ to $F_8$ and for each point adding counts of its two predecessors: $N(\mathbf{B}) = N(\mathbf{B}_L) + N(\mathbf{B}_A) = 35 + 21 = 56$, where $\mathbf{B}_L \equiv \mathbf{B} - <0>$, $\mathbf{B}_A \equiv \mathbf{B} - <1>$. The 8-step path **A..B**, represents binary input string $S_8 = 00101001$.

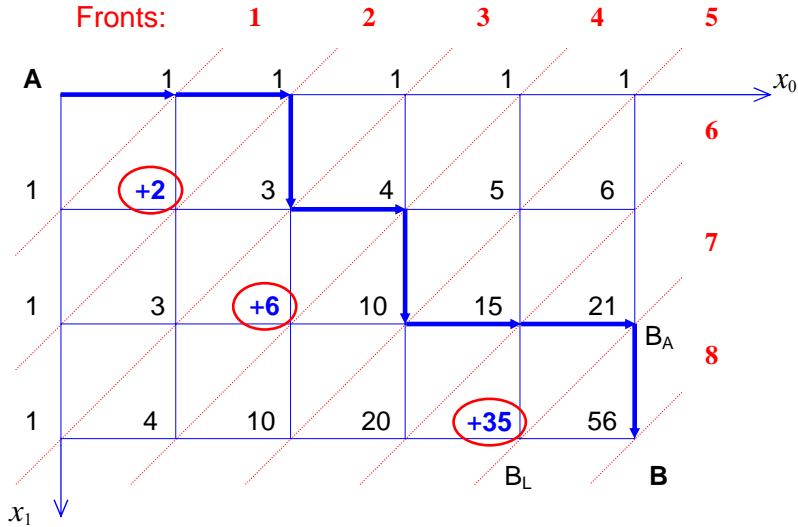

Fig. 1

---

[*] In a more general type of constrained coding[11], the steps $<r>$ allowed from $M$ depend on the entire path $T(M)$ reaching $M$, hence the constraint evaluator is $T(M){:}r$ (requiring extra summations over classes of $T$).
[†] In the conventional EC[1] which uses lex ranking, the building blocks (expressed in our setting) are the counts of paths *from* a given point $M_k$ to a fixed end-point $M_n$ (the quantities $n_S(x_1,x_2,\ldots,x_k)$ in [1] 73).
[‡] Fig 1. is Pascal triangle rotated by 45°. The points with any $x_r < 0$ have path counts 0 (unreachable points via steps $<r>$). AC can also be computed via walks on Pascal triangle with $(x,y)$ 'counts' $p^x q^y$ ([5],[23]).

For unconstrained paths and fixed alphabet $\alpha$, with $M_n \equiv (c_1, c_2,..., c_\alpha)$ (where $c_{1+r}$ is the count of symbol $r$ and $c_1+c_2+...+c_\alpha = n$), the path counts simplify[5] to multinomials:

$$N(M_n) = \binom{c_1 + c_2 + \cdots + c_\alpha}{c_1, c_2, \ldots, c_\alpha} = \frac{(c_1 + c_2 + \cdots + c_\alpha)!}{c_1! c_2! \cdots c_\alpha!} = \frac{n!}{c_1! c_2! \cdots c_\alpha!} \quad (7)$$

***Path index*** is a function $I(T \in E)$ which maps each path $T$ from set $E$ (***enumerative class***) to a unique[*] integer in some interval $D(E) \equiv [D_0, D_1)$ (***index interval***). The uniqueness constrains, via pigeonhole principle[25], the length of $D(E)$ from below as:

$$|D_1 - D_0| \equiv |D(E)| \geq |E| \quad (8)$$

$I(T \in E)$ is called ***compact index*** if $D(E) = [0, |E|)$. The most elemental set $E$, used directly or as a building block for larger classes[†], is the set $E(M)$ (all paths arriving at $M$). We focus now on $E(M)$[‡] and abbreviate $D(M) \equiv D(E(M))$ and $L(M) \equiv |D(M)|$.

Our next aim is to construct a *compact index* $I_i(T_i) \equiv I(T_i \in E(M_i))$ for $i$-step paths $T_i$ from the compact indices $I_{i-1}(T_{i-1}) \equiv I(T_{i-1} \in E(M_{i-1}(r)))$ for $(i-1)$-step paths[§] $T_{i-1}$. We first split the set $E(M_i)$ based on the last step $<r>$ of its paths into $R$ disjoint subsets $A_r$ (where $A_r$ is a set, possibly empty, of all paths arriving at $M_i$ via last step $<r>$):

$$E(M_i) = \bigcup_{r=0}^{R-1} A_r \text{ where } A_r \equiv \{T_i \equiv T_{i-1} + \langle r \rangle : \forall T_{i-1} \in E(M_{i-1}(r))\} \quad (9)$$

The length of *index interval* for paths $T_{i-1}$ in (9) is $L_r \equiv |D(M_{i-1}(r))| = N(M_{i-1}(r))$. To get a compact index $I_i(T_i)$ we choose $L(M_i) \equiv N(M_i)$ which, by denoting $\lambda_r \equiv [M_{i-1}(r):r] \cdot L_r$ and using (5), relates these interval lengths as:

$$L(M_i) \equiv |D(M_i)| = N(M_i) = \sum_{r=0}^{R-1} [M_{i-1}(r) : r] \cdot L_r \equiv \sum_{r=0}^{R-1} \lambda_r \quad (10)$$

Using (10) we construct the index $I_i(T_i)$ by splitting the interval $D(M_i)$ into $R$ disjoint subintervals $\Delta_r \equiv [d_r, d_r + \lambda_r)$ of lengths $\lambda_r$ with $d_0 \equiv 0$, $d_{r+1} \equiv d_r + \lambda_r$, for $r = 0..R-1$, where each $\Delta_r$ indexes the $\lambda_r$ paths $T_i \in A_r$ by reusing the index of $T_{i-1}$ (from eq. (9)) as:

$$I_i(T_i) \equiv I_{i-1}(T_{i-1}) + d_r \quad \text{for} \quad i = 1, 2, \ldots, n \quad (11)$$

$$d_r = \sum_{s=0}^{r-1} \lambda_s = \sum_{s=0}^{r-1} [M_{i-1}(s) : s] \cdot N(M_{i-1}(s)) \quad (12)$$

---

[*] Meaning: $T_1 \neq T_2$ implies $I(T_1) \neq I(T_2)$, $\forall T_1, T_2$. This uniqueness is equivalent to invertiblity of $I(T \in E)$.
[†] One such class is $n$-step front $F_n$ used in **high entropy limit** coding (e.g. mixed radix). They are obtained by adding eq. (5) over all $M_n \in F_n$. A class of *constant entropy*[6,7] segments is useful for composite sources. A class used in [8], of all paths with *fixed sum of symbols*, is an improper (at best approximate) $E$-class.
[‡] Since $E(M_0)$ is an empty set $\varnothing$, for notational convenience we define index function $I_0(T_0) \equiv I(T \in \varnothing) \equiv 0$.
[§] Note that for all 1-step points $M_1$ there is a unique compact index function $I_1(T_1) = I(T_1 \in E(M_1)) = 0$.

In words, the $i$-step paths $T_i$ from set $A_r$ are indexed by offsetting index of their $(i-1)$-step prefixes $T_{i-1}$ by $d_r$, where $d_r$ simply keeps track of the interval length used up so far for the intervals $\Delta_s$ which are to the left of $\Delta_r$ (i.e. $s<r$). Expressing paths in (11) in terms of their steps $<r_j>$ ($j=1..i$, and $r_i \equiv r$) as $T_i \equiv r_1 r_2 ... r_i$ and $T_{i-1} \equiv r_1 r_2 ... r_{i-1}$, we obtain:

$$I_i(r_1 r_2 \ldots r_i) = I_{i-1}(r_1 r_2 \ldots r_{i-1}) + d_r \tag{13}$$

which, in terms of sequence $S_n = a_1 a_2 \ldots a_n$ and upon expanding recurrence, becomes:

$$I_i(a_1 a_2 \ldots a_i) = I_{i-1}(a_1 a_2 \ldots a_{i-1}) + \sum_{r=0}^{a_i-1} [M_{i-1}(r):r] \cdot N(M_{i-1}(r)) \tag{14}$$

$$I_n(a_1 a_2 \ldots a_n) = \sum_{i=1}^{n} \sum_{r=0}^{a_i-1} [M_{i-1}(r):r] \cdot N(M_{i-1}(r)) \tag{15}$$

Eqs. (5),(15) represent a finer resolution form of the EC indexing formula (Prop. 2 in [1], p. 74). The essential difference between (15) and (P2) of [1] is that (P2) melds the two factors forming individual terms of (15) into a single *black box* quantity $n_S(x_1, x_2, \ldots x_i)$, taken as an opaque input to EC and not analyzed further. In contrast, we separate the constraints $[M:r]$ factors from the counts $N(M)$ factors in the index formula (15) and then find further important *general* relations (5) for the counts factor which are completely absent (even as a question) in [1]. This absence is the principal reason for the "*major difficulty*" (cf. [18] p. 49) in solving the precision problem of EC since the optimal EC quantization resides precisely in the overlooked relations (5).

For unconstrained binary source, eqs. (5), (15) (using $(x,y) \equiv (x_0, x_1)$, $k \equiv y$, $n \equiv x+y$, input $S_n = a_1 a_2 \ldots a_n$, path points $M_i \equiv (x_i, y_i)$, $n_j \equiv$ 0-based offset of $j$-th 1 in $S_n$) simplify to:

$$N(x,y) = N(x-1,y) + N(x,y-1) = \binom{x+y}{y} \equiv \binom{n}{k} \equiv C(n,k) \tag{16}$$

$$I_n(a_1 a_2 \cdots a_n) = \sum_{i=1}^{n} [a_i > 0] \cdot N(x_i - 1, y_i) = \sum_{j=1}^{k} \binom{n_j}{j} \tag{17}$$

Our eq. (17) corresponds to eq. (4) in [1], where the $n_j$ of (17) is replaced by $(n-n_j)$. This difference illustrates one advantage of colex ([26] p. 59, "*Colex superiority principle*") over lex indexing used in [1]: at each step $j=1..k$, the addend in (17) depends only on the string processed so far, while the addends in eq. [1].(4) depend also on the size $n$ of the entire string. Hence, the colex coder can easily terminate the coding loop based on a variety of termination conditions encountered in the first pass (e.g. for VF or VV coding, where the sum value accumulated so far may signal termination[6,7]), while the lex de/coder must know in advance the block size $n$ (usable mainly for FV coding).

A numeric example of indexing via (17) for input $S_8 = 00101001$ is shown in Fig. 1. The encircled addends $N(x_i-1, y_i)$ are the path counts of the *left neighbors* at the end of each vertical step. The index of $S_8$ is: $I(S_8) = C(2,1) + C(4,2) + C(7,3) = 2 + 6 + 35 = 43$.

# Quantized Indexing

The impracticality of exact EC and the need for its quantization (reduced precision) is evident even for the simplest binary coding via (17). The binomial coefficients used as addends in (17) require arithmetic precision of the size of output (the entropy $H(S_n)$), which is generally $O(n)$ bits. Since the number of operations in (17) is $O(n)$ adds of $O(n)$ bit numbers, the coder needs $O(n^2)$ bit operations[10]. The addends are also impractical to compute on the fly, so they need to be precomputed, e.g. for some maximum block size $n$ into a table holding $\sim n^2/4$ entries of size $O(n)$ bits/entry, resulting in table sizes of $O(n^3)$ bits. For an alphabet of size $\alpha$, the arithmetic complexity is $O(n^2 \log(\alpha))$ bit operations, while the table sizes are exponential in $\alpha$ as $O(n^{\alpha+1})$ bits.

Our quantization tool is a ***sliding window integer*** (SWI), defined as the integer function $Q = SW(g, w, s)$, where $g$ *significant bits* of $Q$ are given as integer *mantissa* (window) $w$, while the *exponent* (shift) $s$ specifies the number of zeros in the less significant bits of $Q$. Although the abstract SWI appears *structurally* identical to abstract floating point numbers (FP), the operators, concrete structures and application controls differ between the two types of numbers. SWI is a hybrid, combining functionality of unlimited size integers with some structural constraints of FP. Unlike FP, the arithmetic with SWI operands maintains the arithmetic properties of integers, such as associativity of additions[*]. The essential structural and functional properties of SWI are as follows:

$$Q = SW(g, w, s) = w \cdot 2^s \quad \text{and} \quad (s > 0) \Rightarrow 2^{g-1} \leq w < 2^g \qquad (18)$$

$$SW(g, w, s) = \overbrace{1xx\cdots x}^{g-\text{bit } w}\,\overbrace{0000\cdots 0000}^{s \text{ bits}} \quad \text{for } (s > 0) \qquad (19)$$

i) **SW *arithmetic*** (in any combinations of SWI and integers) yields exactly the same result as the corresponding integers $Q$ in the expanded form (18) would produce. The sole difference is that SWI operands generally require $(s+g)/g$ times fewer bit operations than integer operands of the same magnitude, $(s+g)$ bits.

ii) **SW *rounding*** $Q \equiv \lceil X \rceil_{sw}$ maps an integer $X$ to the *nearest* SW integer $Q \geq X$ [†]. Unlike FP, due to (i) the SW rounding functions as a ***delayed rounding***[‡].

Since the source of EC precision problem are counts $N(M)$ used in the index (15), the "obvious idea"[1,18] would be to quantize $N(M)$ via $L(M) \equiv \lceil N(M) \rceil_{sw}$. While such $L(M)$ does satisfy pigeonhole principle (8), $L(M) \geq N(M)$, this kind of global, one-step majorization of all $N(M)$ cannot guarantee validity of the EC interval composition underlying the index recurrence (11). Namely, note that the set decomposition $E(M_i) = \bigcup_r A_r$ in (9) implies that the index interval $D(M_i)$ of set $E(M_i)$ must contain all disjoint intervals $\Delta_r$ used as index intervals for sets $A_r$ in (11), which then constrains the lengths of these intervals, $\lambda_r \equiv |D(A_r)| = [M_{i-1}(r):r] \cdot L(M_{i-1}(r))$, $r = 0..R-1$ and $L(M_i) \equiv |D(M_i)|$, as:

$$L(M_i) \geq \sum_{r=0}^{R-1} \lambda_r \equiv \sum_{r=0}^{R-1} [M_{i-1}(r):r] \cdot L(M_{i-1}(r)) \qquad (20)$$

---

[*] Hence, the conventional FP, in hardware or in libraries, is not suitable for SWI implementations.
[†] And similarly for the rounding down operator: $Q \equiv \lfloor X \rfloor_{sw}$ yielding the nearest $Q \leq X$.
[‡] For example, $\lceil X+Y+Z \rceil_{sw}$ is equivalent via (i) to computing integer $X+Y+Z$ in full precision (albeit quicker) and then rounding only the *final result* up, while the FP evaluation of $X+Y+Z$ would also round each partial sum, which will generally be larger than $\lceil X+Y+Z \rceil_{sw}$.

If we were to just set *globally* $L(M) = \lceil N(M) \rceil$ for all $M$ and for some quantization $\lceil X \rceil$, recalling that via (5) the non-quantized values $N(M)$ satisfy exact equality in (20), then this quantization must satisfy e.g. $\lceil Z \rceil \geq \lceil X \rceil + \lceil Y \rceil$ for arbitrary non-quantized $Z = X + Y$. Any simple rounding up (including SW rounding) *"will not work at all"* [*], as Rissanen noted[18]. In our EC formulation, the solution suggests itself naturally: we *propagate* the three constraints, the SW rounding (combining the pigeonhole principle (8) and the $g$-bit precision constraints) and the interval composition constraint (20), from $F_1$, where they already hold, to $F_2$, $F_3$... *front by front*. Thus QI computes $L(M)$ by replacing (5) with:

$$L(M_i) = \left\lceil \sum_{r=0}^{R-1} [M_{i-1}(r):r] \cdot L(M_{i-1}(r)) \right\rceil_{SW}, \quad L(M_0) \equiv 1, \; i = 1,2,\ldots \qquad (21)$$

Since $L(M_0) = N(M_0)$, (5) and (21) imply (via front by front induction) $L(M) \geq N(M)$ for any $M$, hence (8) is satisfied globally. SW rounding used in (21) implies (by its definition (ii)) that the resulting $L(M_i)$ is the *smallest* interval length satisfying (20) for any given precision $g$ for $L(M_i)$. Using (20) instead of (10), we can replicate the index construction in (11), with the only differences that now we do not assume $L(M) = N(M)$ (it was used for the lengths $\lambda_r$, thus we keep $L(M)$ in (12)) and that the intervals $\Delta_r$ may not cover the entire interval $D(M_i)$, hence the index has some redundancy. The simple notation changes of (11) into (14)-(15) follow through as before, yielding the final QI index equations:

$$I_i(a_1 a_2 \ldots a_i) = I_{i-1}(a_1 a_2 \ldots a_{i-1}) + \sum_{r=0}^{a_i - 1} [M_{i-1}(r):r] \cdot L(M_{i-1}(r)) \qquad (22)$$

$$I_n(a_1 a_2 \ldots a_n) = \sum_{i=1}^{n} \sum_{r=0}^{a_i - 1} [M_{i-1}(r):r] \cdot L(M_{i-1}(r)) \qquad (23)$$

To obtain QI redundancy $\delta(g)$ due to SW rounding in (21), note that for any $X$, whenever $Q = \lceil X \rceil_{SW}$ yields $Q > X$, the $w$ normalization in (18) implies $w \geq 2^{g-1}$, hence the expansion factor $f \equiv Q/X$ is bound as $f < (2^{g-1}+1)/2^{g-1}$. Thus $\delta(g) < \log(1 + 1/2^{g-1}) < \log(e)/2^{g-1}$ bits/sym[†]. To keep the total output excess for $n$ symbols below $c$ bits, the SW precision $g$ must be at least $1 + \log(\log(e)) + \log(n/c) \approx 1.53 + \log(n/c)$. Since the unused index values of intervals $D(M)$ are known from the common $L(M)$ tables, these **gaps can be reclaimed** [‡] by the modeling engine as Esc code space (of size adjustable via $g$) or by the coder itself for input segmentation, QI *frugal bits*[22] mode, output structuring, steganography,... etc.

---

[*] E.g. consider ceiling function for $X=1.1$, $Y=1.2$: $\lceil Z \rceil = \lceil 1.1 + 1.2 \rceil = \lceil 2.3 \rceil = 3 < \lceil 1.1 \rceil + \lceil 1.2 \rceil = 2 + 2 = 4$.
[†] Measured values (for the entire $L(M)$ tables) of max excess $n \cdot \delta$ for $\alpha=2$, $g = \lceil \log(n) \rceil + 1$ and $n \leq 2^{14}$ are below 0.7 bits (avg 0.3), while the theoretical bound is 1.44 bits. QI coder normally uses $g = 32$ and $n \leq 2^{12}$.
[‡] AC, lacking the exact addend tables, cannot as easily or as efficiently reuse these gaps in its code space.

# Implementation Notes and Test Results

**N1.** The index (23) can be computed in *forward* ($i=1..n$) or in *reverse* ($i=n..1$) direction, which functions exactly as **LIFO** and **FIFO** coding modes[19] of AC. LIFO mode is faster, simpler and has lower redundancy while FIFO mode allows decoder to start as soon as the initial $g$ bits were received (encoder may output data in both modes with a $g$ bit delay). The *carry problems* and solutions in FIFO mode are the same as with AC[18-20,24].

**N2.** The addends $L(M)$ are precomputed via (21) into tables which may cover all points to some **n** (with symmetry reductions) or *constant entropy* V-shaped[6,7] tables or long *low entropy* strips near lattice axes. To *save table space*, one may keep only *every m-th front* (reducing table sizes by factor *m*) and use recurrences (21) (except that here it is usually *better not to round* at all these interpolated $L(M)$[24]) to compute the omitted fronts on the fly[24]. The table values can also be reduced from the full ($w, s$) content, since the *exponents* s are very regular and can be quickly computed or *interpolated* from much smaller tables. For example, QI coder uses $g=32$ and keeps only mantissas. It computes on the fly the exponents for binomials $n!/[k!(n-k)!]$ from an array of $n$ entries ($\log(k!)$ in fixed point), freeing thus the space of $n^2/4$ exponents at a negligible speed cost.

**N3.** The **high entropy limit** codes[23] use fronts $F_i$ as enumerative classes $E$, thus the tables grow linearly in $n$. Instead of $n^2/4$ binomials of (16), the tables have only $n$ entries, $R^i$ for fixed radix $R$, $i!$ for *permutations*, Catalan numbers $C_i$ for *trees*,... etc[23]. The QI versions of **mixed radix** $R_i$ tables are[24]: $L_i = \lceil R_i \cdot L_{i-1} \rceil_{sw}$, $L_0 = 1$ (for permutations: $L_i = \lceil i \cdot L_{i-1} \rceil_{sw}$).

**N4.** QI coder uses mixed radix codes [N3] to **eliminate bit fraction losses** at the block boundaries by using for $R_i$ the leading 16 bits of value $L(M_n)$, common to all paths $T(M_n)$. It encodes the leading 16 bits of the computed index $I_n(T(M_n))$ as a digit in this radix.

**N5.** For **multi-alphabet coding**, the multinomial tables (7) are not practical due to table sizes $\sim n^{\alpha+1}$. QI uses factorization of multinomials into products of binomials (guided by the alphabet prefix codes), which allows the reuse of quantized binomial tables for $\alpha > 2$, with the resulting algorithm[24] resembling the newer $\log(\alpha)$ complexity variants of AC[22].

**N6.** The *encoding of symbol counts* can use a streamlined form of multi-alphabet QI (optimized for binomial or hypergeometric distribution of counts per block). For just a few blocks or for not overly sparse inputs, the counts can be coded quicker and nearly as well using MTF or precomputed (e.g. for binomial distribution) Huffman codes.

**N7.** The *modeling* paradigms differ substantially[23] between EC/QI (Kolmogorov, finite sequence and its exact properties, counts, BWT,...) and AC (Shannon, infinite sequence "source" and its limit averages, probabilities, PPM,...), although each coder can *in principle* work both ways[21,23] (with penalties in 'non-native' mode). E.g. QI can code the AC way from the 'next symbol probabilities' $p(a)$ using *lattice jumps*, but these require mantissa scaling on each jump, reducing the QI speed to that of AC. In native coding mode, but with only AC modeling engine available, QI separates *output* streams for each probability class, thus it codes quickly again and nearly optimally. In the native QI modeling[23] the *input* is segmented into enumerative classes (e.g. via BWT, grammars).

Below are few *test results* of QI vs. AC ('best' settings[22]) for input sizes N and given # of 1's (binary order 0 coders; K≡$2^{10}$ bits; 500 random inputs/result; columns Speed: coding times *ratio* A/Q, N: output *size %* (A/Q-1)·100; array Vary ≡ int32 {...,-2,-1,0,+1,+2,...}).

| #1's | N: 4K | Speed | N: 8K | Speed | N: 32K | Speed | N: 128K | Speed |
|---|---|---|---|---|---|---|---|---|
| 8    | 6.846    | 68.3 | 6.421  | 112.8 | 5.447  | 199.6 | 5.966  | 247.5 x |
| 16   | 4.175    | 59.7 | 3.830  | 78.5  | 3.389  | 138.1 | 3.730  | 168.0 |
| 32   | 2.297    | 49.7 | 2.090  | 58.9  | 2.096  | 95.9  | 2.220  | 117.2 |
| N/64 | 1.370    | 40.3 | 0.606  | 41.0  | 0.186  | 41.7  | 0.073  | 42.5 |
| N/32 | 0.897    | 30.8 | 0.343  | 33.7  | 0.123  | 34.2  | 0.049  | 34.5 |
| N/16 | 0.505    | 21.8 | 0.197  | 25.3  | 0.084  | 24.6  | 0.040  | 24.8 |
| N/8  | 0.359    | 14.4 | 0.155  | 16.7  | 0.069  | 16.8  | 0.045  | 16.8 |
| N/4  | 0.288    | 9.2  | 0.138  | 10.8  | 0.083  | 10.6  | 0.068  | 10.5 |
| N/2  | 0.509    | 6.6  | 0.445  | 6.6   | 0.367  | 6.4   | 0.332  | 6.4 |
| Vary | 110.899 %| 21.9 | 96.736 | 19.6  | 71.308 | 16.5  | 52.580 | 14.1 |